\documentclass[10pt]{article}

\usepackage{amsmath}
\usepackage{amssymb}

\usepackage{graphicx}

\usepackage{cite}

\usepackage{color} 

\usepackage[labelfont=bf,labelsep=period,justification=raggedright]{caption}

\makeatletter
\renewcommand{\@biblabel}[1]{\quad#1.}
\makeatother

\date{}

\begin{document}

\begin{flushleft}
{\Large
\textbf{Mean field approximation for biased diffusion \textcolor{black}{on Japanese} inter-firm trading network}
}
\\
 Hayafumi Watanabe $^{1,\ast}$
\\
\bf{1} Hottolink Inc., Chiyoda-ku, Tokyo, Japan
\\
$\ast$ E-mail:  h.watanabe@hottolink.co.jp
\end{flushleft}

\section*{Abstract}
\textcolor{black}{By analysing the financial data of firms across Japan, a nonlinear power law with an exponent of 1.3 was observed between the number of business partners (i.e. the degree of the inter-firm trading network) and sales. In a previous study using numerical simulations, we found that this scaling can be explained by both the money-transport model, where a firm (i.e. customer) distributes money to its out-edges (suppliers) in proportion to the in-degree of destinations, and by the correlations among the Japanese inter-firm trading network. However, in this previous study, we could not specifically identify what types of structure properties (or correlations) of the network determine the 1.3 exponent. In the present study, we more clearly elucidate the relationship between this nonlinear scaling and the network structure by applying mean-field approximation of the diffusion in a complex network to this money-transport model. Using theoretical analysis, we obtained the mean-field solution of the model and found that, in the case of the Japanese firms, the scaling exponent of 1.3 can be determined from the power law of the average degree of the nearest neighbours of the network with an exponent of -0.7.} 
%
\section*{Introduction}
Complex networks have been \textcolor{black}{extensively studied over the last decade} \cite{Newmanreview, boccaletti2006complex,Brabashreview}.
Problems of transport on complex networks, which are some of \textcolor{black}{the} most fundamental problems \textcolor{black}{concerning} the physics of complex networks, have \textcolor{black}{also} been \textcolor{black}{intensively studied}. 
For example, random walks on complex networks have been investigated from various viewpoints \cite{pagerank,random_complex, page_average}. 
One application of transport on complex networks is PageRank, which corresponds to the steady-state density of transport caused by random walks on the Internet. PageRank is one of the most successful indices evaluating the importance of web pages and has been utilized by internet search engines.  \par
\textcolor{black}{Most} studies regarding transport on complex networks have been based on theoretical approaches; \textcolor{black}{however}, recently the problems of actual transport on complex networks have also \textcolor{black}{been} studied \textcolor{black}{by} using massive data analysis. Examples of such problems include, airport traffic on the worldwide airport network\cite{hikouki}, the number of trains on the Indian railway network\cite{IndianRailWay}, the flow of viewers on portal sites\cite{browsing} and control flows on stock-ownership networks\cite{kabu1,kabu2,kabu3}. \par
\textcolor{black}{In a previous study, we analysed the empirical data from an inter-firm trading network, which consisted of approximately one million Japanese firms, and the sales of these firms (a sale corresponds to the total in-flow into a node) to investigate the actual transport phenomenon in a complex network \cite{Watanabe2011}.
This inter-firm trading network is known to be a typical complex network with a power law degree distribution \cite{Ohnishi2009,Watanabe2011}, a negative degree-degree correlation\cite{Watanabe2011, Tamura2012}, a small world property\cite{Ohnishi2009}, community structures \cite{Iino2010}, power laws of money flows \cite{Tamura2012,Tamura2012b} and an asymmetric behaviour of authorities and hubs explained by a network-evolution model based on the preferential attachment rule \cite{Miura2012,Miura2012b}. 
To be more precise, we obtained the following results from Ref. \cite{Watanabe2011}: 
(i) we found a non-trivial empirical power law with an exponent of 1.3 between the number of business partners in a firm and its sales by analysing the data; (ii) we introduced a simple money-transport model in which a firm (i.e. a customer) distributes money to its out-edges (i.e. suppliers) in proportion to the in-degree of the destinations; (iii) using numerical simulations, we found that the steady flow of the abovementioned inter-firm trading network derived from this model can approximately reproduce a power-law scaling with an exponent of 1.3 between the number of business partners (i.e. degrees of the network) and sales. Furthermore, the sales distribution of the actual firms obeys the power law distribution with an exponent of approximately 1. Moreover, the sales of individual firms derived from the money-transport model are shown to be proportional to the real sales on the average. However, we also showed that the simple random walk model (i.e., PageRank) in which a firm is assumed to evenly distribute to all its outgoing neighbours does not reproduce the empirical scaling with an exponent of 1.3. This result implies that PageRank does not correspond to the sales. }
Note that scaling with an exponent of 1.3 was observed are observed  \textcolor{black}{by analysing the abovementioned Japanese financial data} not only between the sales and degrees but also between other important firm-size indicators, for instance, between sales and number of employees, between profits and number of employees, between profits and degrees, etc. \cite{watanabe2012relations}.
 \par
\textcolor{black}{In the previous study \cite{Watanabe2011}, we could not specifically identify what types of structure properties (or correlations) of the Japanese inter-firm trading network determine the 1.3 exponent, although we found that we need a particular network structure to be reproduced in our framework using numerical simulations. Therefore, the current study is devoted to the theoretical study of the results of Ref. \cite{Watanabe2011}.
In particular, we apply a mean-field approximation to the models to clarify how the exponents of the power-law relationships, specifically between the degree of the inter-firm trading network (i.e. the number of business partners) and sales depend on the network structure. }
\par
In this paper, we start \textcolor{black}{in Section 2} by reviewing the models and the numerical results of the previous study.
Next, we introduce the mean-field approximation for these models and reveal the relationships between the power-law exponent, models, and networks \textcolor{black}{in Section 3.} Finally, we summarize and \textcolor{black}{present our} conclusions in \textcolor{black}{Section 4.}
\section*{Power laws for diffusion \textcolor{black}{on inter-firm} trading network}
In this section, we review the results of Ref. \cite{Watanabe2011}.
\textcolor{black}{To understand the \textcolor{black}{empirical scaling with exponent 1.3} between degrees and sales,} 
we introduced the following money-transport models \cite{Watanabe2011}. 
\\
Model-1 (Equi-partition model):
\begin{equation}
x^{}_{m}(t+1)=\sum^{N}_{i=1}A_{im}\frac{1}{k_i^{(out)}}x^{}_{i}(t), \label{model1}
\end{equation}
Model-2 (Weighted partition model):
\begin{equation}
x^{}_{m}(t+1)=\sum^{N}_{i=1}A_{im}\frac{k_m^{(in)}}{\sum^{N}_{j=1}A_{ij}k_{j}^{(in)}}x^{}_{i}(t), \label{model2}
\end{equation}
where $x_m(t)$ denotes the \textcolor{black}{sales} of \textcolor{black}{node $m$} at time step $t$, $A_{ij}$ is an adjacency matrix, $k_m^{(in)}$ is the in-degree of \textcolor{black}{node $m$}, $k_m^{(out)}$ is the out-degree of \textcolor{black}{node $m$} and $N$ is the number of nodes on the network.
Note that for $k_i^{(out)}=0$ in Eq. (\ref{model1}) or $\sum^{N}_{j=1}A_{ij}k_{j}^{(in)}=0$ in Eq. (\ref{model2}), we omit the contributions of the \textcolor{black}{node $i$}. \par
In Model-1, a node representing a business firm is assumed to \textcolor{black}{evenly} distribute its total in-flow (i.e. sales) at time t to all its outgoing neighbours in the next time step. This transport is equivalent to the simple probability diffusion of the PageRank model in the case of no random \textcolor{black}{spontaneous jumping \cite{pagerank}.} \textcolor{black}{However, for} Model-2, a node distributes its sales to its outgoing neighbours in proportion to the destinations' in-degrees. This model is one of the so-called biased random walk models \cite{bias}. \par 
We used two types of networks for numerical experiments to clarify the dependencies on the firm network \cite{Watanabe2011}. 
\textcolor{black}{The data set used in the generation of networks and the data analysis in Refs. \cite{Watanabe2011,watanabe2012relations} was provided by Tokyo Shoko Research, Ltd. and contains approximately one million firms, which practically covers all active firms in Japan. For each firm, the data set contains the annual sales and a list of business partners in 2005, categorized into suppliers and customers \cite{Watanabe2011, Ohnishi2009,watanabe2012relations}. } \par
The first network \textcolor{black}{is} the real firm network whose nodes \textcolor{black}{are firms} and \textcolor{black}{whose edges are} defined by the following rule: if \textcolor{black}{firm} $i$ purchases goods and/or services from \textcolor{black}{firm $j$}, or, equivalently, if money flows \textcolor{black}{from firm $i$ to firm $j$}, we connect node $i$ to node $j$ with a directed link \textcolor{black}{(there are 961,318 nodes and 3,783,711 edges)} \cite{Ohnishi2009,Watanabe2011}. \textcolor{black}{This network was generated by using the business partners data in the abovementioned data set.}
\textcolor{black}{The firm network is a typical complex network whose main properties are as follows}: (i)\textcolor{black}{a} power-law degree distribution with exponent 1.3, (ii)\textcolor{black}{a} negative degree-degree correlation and (iii)\textcolor{black}{a} small-world property (\textcolor{black}{the mean distance} between nodes is 5.62 and the maximum distance is 21) \cite{Ohnishi2009,Watanabe2011,Iino2010}. \par
The second network \textcolor{black}{is} the shuffled firm network, \textcolor{black}{which is} an almost-uncorrelated network with the same degree distribution as the firm network \textcolor{black}{generated by the Maslov-Sneppen algorithm \cite{shuffle,shuffle2}}.  Note that we simulated the time evolution of the models on the largest strongly connected component (LSCC) for each network to neglect boundary effects. 
\textcolor{black}{Then, \textcolor{black}{total money is conserved},  $\sum^{N}_{i=1}x_i(t)=\sum^{N}_{i=1}x_i(1)$ on the LSCCs of both networks.} \par
For all combinations of models and networks, we obtained \textcolor{black}{by numerical simulations} the following power law for large $k^{(in)}$ ($t>>1$) \cite{Watanabe2011}:
\begin{equation}
x(k^{(in)}) \propto \left\{ k^{(in)} \right\}^{\beta}. 
\end{equation}
Here, $x(k^{(in)})$ is the conditional mean of $x$ for \textcolor{black}{a} given $k^{(in)}$ as a function of $k^{(in)}$ \textcolor{black}{and is} defined by
\begin{equation}
x(k^{(in)})=\frac{\sum_{l \in \left\{l|d(i) \leq k_l^{(in)} < d(i+1) \right\}}x_l}{\sum_{l \in \left\{l|d(i) \leq k_l^{(in)} < d(i+1) \right\}}1}, 
\end{equation}
where $d(i)$ is the separator of \textcolor{black}{box $i$}, taken evenly in logarithmic space, e.g. $d(i)=2^i \quad (i=0,1,2,\cdots)$, and $k^{(in)}$ on the right-hand side is represented by the geometric mean $\sqrt{d(i) \cdot d(i+1)}$. 
The exponent $\beta$ depends on both the network and model. These dependencies are summarized in Table 1. Note that Model 2 for the firm network reproduces the empirical scaling with exponent 1.3.
%
\section*{Mean-field approximation for models}
In this section, we apply the mean-field approximation to the models to understand the relationship between the exponents, models, and network structures as shown in Table 1.
Here we neglect the directions of the networks for simplicity (\textcolor{black}{scalings nearly identical to} results presented in \textcolor{black}{the} previous section can be obtained by numerical simulation, regardless of whether \textcolor{black}{or not} the edge directions are neglected).  
\textcolor{black}{By assuming $\sum_{i=1}^{N}x_i(1)=1$, we can regard the models as \textcolor{black}{Markov processes} with an existing probability of $x_M(t)$,
\begin{equation}
 x_{M}(t+1)=\sum_{i=1}^{N}Q_{Mi} \cdot x_i(t), \label{malkov}
 \end{equation} 
 where $Q_{Mi}(t)$ is the transition probability from \textcolor{black}{node $i$ to node $m$} \cite{Watanabe2011}.
 } 
Note that \textcolor{black}{although} we neglect the direction of edges in this paper, the following discussion can be extended to the case of a directed network by adding some assumptions and calculations. \par
\par 
In general, one \textcolor{black}{of the} mean-field solutions of the Markov process defined by \textcolor{black}{the} transition matrix
\begin{equation}
Q_{Mi}=\frac{A_{iM} \cdot g(k_M)}{\sum_{r=1}^{N}A_{ir} \cdot g(k_r)} \label{qmi}
\end{equation}
is given \textcolor{black}{in Ref. \cite{baronchelli2010mean}} by
\begin{equation}
R(k)=\frac{kP_{k}(k)g(k)\sum_{k'}g(k')P_{k'|k}(k'|k)}{\sum_{q}\sum_{k'}qP_{k}(q)g(q)g(k')P_{k'|k}(k'|q)}, \label{kinji}
\end{equation}
where \textcolor{black}{$g(k)$ is the weight of the transition probability to a node with the degree $k$, which is defined as a function of the degree $k$}, $R(k)$ is the probability that the walker is at any node of degree $k$ in the steady state, $P_k(k)$ is a probability density function of \textcolor{black}{degree $k$} and $P_{k'|k}(k'|k)$ is the conditional probability that a node of degree $k$ is connected to a node of degree $k'$.  
To obtain this solution, \textcolor{black}{the authors of Ref. \cite{baronchelli2010mean} } employed mainly two approximations. 
First, \textcolor{black}{they used the annealed network approximation where they regarded the sets of firms with common degrees in the original network as nodes of the approximated network, and the weighted edges of the approximated network are associated with a probability that two nodes of degrees $k$ and $k'$ are connected, }   
\begin{equation}
\bar{A}(k,k') = \frac{1}{NP_{k}(k)}\frac{1}{NP_{k}(k')}\sum_{i \in k}\sum_{j \in k'}A_{ij},  \label{akk}
\end{equation}
where $i \in k$ is denoted as a sum over the set of nodes of degree $k$ \cite{baronchelli2010mean}.
Second, \textcolor{black}{they} replaced $Q_{Mi}$ \textcolor{black}{with average} probability of an interaction between nodes \textcolor{black}{of degree $k$} and $k'$, defined by  
\begin{equation}
\bar{Q}(k', k)=\frac{[NP_k(k)]^{-1}\sum_{i \in k}\sum_{j \in k'}g(k_j)A_{ij}}{[NP_k(k)]^{-1}\sum_{i \in k}\sum_{r}g(k_r)A_{ir}}. \label{qkk}
\end{equation}
\textcolor{black}{Then, we can replaced the Markov process given by Eq. \ref{malkov} with 
\begin{equation}
R(k;t+1)=R(k;t)-\sum_{q} \bar{Q}(q,k)  R(k;t)+\sum_{q}\bar{Q}(k,q)  R(q;t), \label{time}
\end{equation}
where $R(k;t)$ is the probability that the walker is at any node of degree $k$ at time $t$. }
\textcolor{black}{By substituting Eq. \ref{kinji} into this equation and by using the degree detailed balance condition, $k'P_k(k')P_{k'|k}(k'|k)=kP_k(k)P_{k'|k}(k|k')$, \textcolor{black}{we confirm} that Eq. \ref{kinji} is a steady state of Eq. \ref{time}. }
More details on these approximations are available in Ref. \cite{baronchelli2010mean}  \par  
 Next, we apply Eq. \ref{kinji} to our models given by Eqs. \ref{model1} and \ref{model2}. 
In Model-1, the transition probability is uniform\textcolor{black}{;} that is, $g(k)=1$ in Eq. \ref{qmi}.
Therefore, the  conditional mean of sales $x(k)$ for the given degree $k$ (i.e. $R(k)$ per node), is written as 
\begin{equation} 
x(k)=\frac{R(k)}{NP_k(k)}=\frac{k \cdot 1 \cdot P_{k}(k) \cdot 1}{\sum_{q}q \cdot 1 \cdot P_k(q) \cdot 1}\cdot \frac{1}{NP_{k}(k)} \propto k. \label{knn1}
\end{equation}
This scaling agrees with simulation results for Model-1 for the firm network and the shuffled network. \par
Similarly, by applying Eq. \ref{kinji} \textcolor{black}{to} the case that $g(k)=k$ in Eq. \ref{qmi} (corresponding to Model-2), we obtain
\begin{equation}
R(k)=\frac{k^{2}P_{k}(k)k_{nn}(k)}{\sum_{q}q^2P_k(q)k_{nn}(q)}, \label{rkk}
\end{equation}
where we denote the average degree of nearest \textcolor{black}{neighbours} as 
$k_{nn}(k) \equiv \sum_{q}qP_{k'|k}(q|k)$ \cite{knn}. 
Thus, we find 
\begin{equation}
x(k)=\frac{k^{2}P_{k}(k)k_{nn}(k)}{\sum_{q}q^2P(q)k_{nn}(q)}\cdot\frac{1}{NP_k(k)} \propto k^{2}k_{nn}(k). \label{knn2}
\end{equation}
From Fig. 1(a) in Ref. \cite{Watanabe2011}, \textcolor{black}{we see} that $k_{nn}(k) \propto k^{-0.7}$ for a large degree $k$ in the case of the firm network. 
We substitute this empirical result into Eq. \ref{knn2} to obtain
\begin{equation}
x(k) \propto k^{2}k_{nn}(k) \propto k^{1.3}. \label{k13} 
\end{equation}
This scaling corresponds to the simulation results \textcolor{black}{for Model-2} for the firm network and \textcolor{black}{empirical observation.}
Note that Eq. \ref{knn2} implies that the exponents $\beta$ depend on the average degree of nearest neighbourhoods, $k_{nn}(k)$ for \textcolor{black}{Model-2.} \par
\textcolor{black}{For} the shuffled network for Model-2, which is the uncorrelated network shown in Fig. 1(a) of Ref. \cite{Watanabe2011} (i.e. $k_{nn}(k) = const.$), we obtain the following from Eq. \ref{knn2}:  
\begin{equation}
x(k) \propto k^2 \cdot 1=k^2. \label{knn3} 
\end{equation} 
This equation agrees with the numerical result. \par
\begin{figure}[!ht]
\centering
\includegraphics[width=8.5cm]{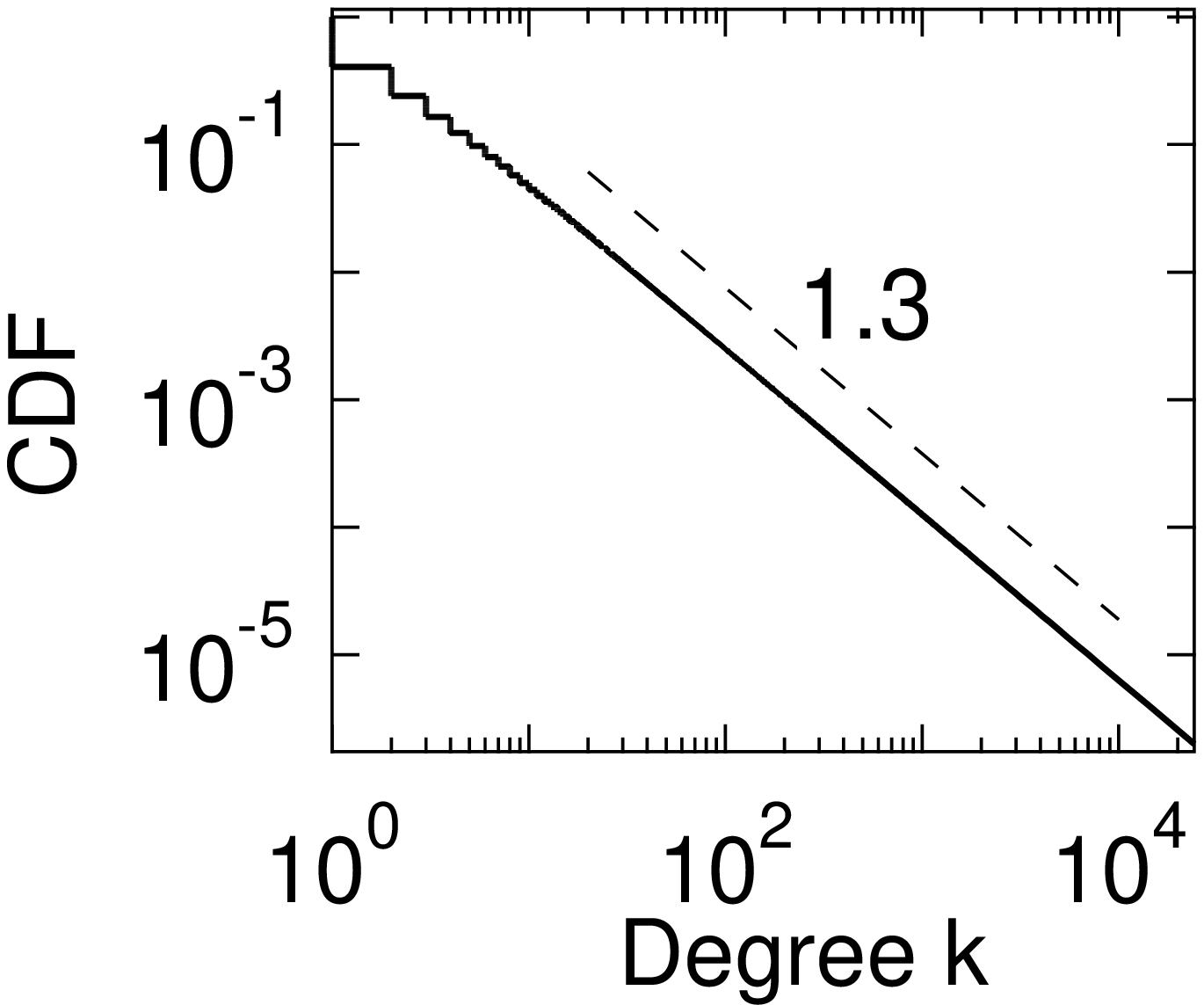}
\caption{\textcolor{black}{{\bf The cumulative distribution function of degree of the artificial networks.}  We confirm that degree obeys the power law distribution with exponent -1.3, namely, $P_k(k) \propto k^{-2.3}$.}}
\label{fig1}
\end{figure}
\begin{figure}[!ht]
\centering
\includegraphics[width=8.5cm]{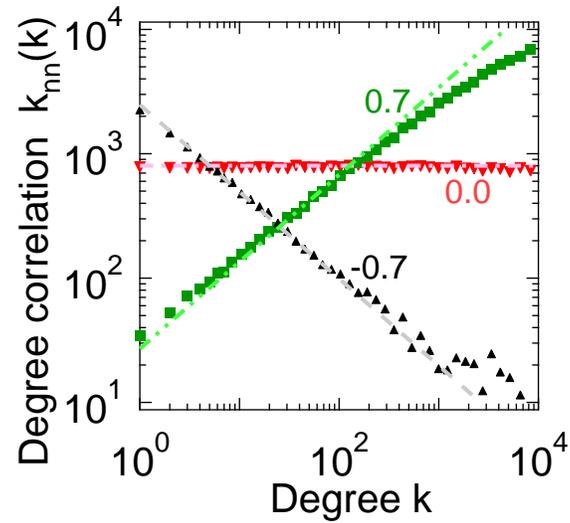}
\caption{\textcolor{black}{{\bf The averages degree of nearest neighbours as a function of degree of the artificial networks, $k_{nn}(k)$.} The black triangles indicate the network for $\gamma=-0.7$, the red nablas for $\gamma=0.0$ and the green squares for $\gamma=0.7$. The black dashed line is proportional to $k^{-0.7}$, the red dash-dotted line is proportional to $k^{0.0}$ and the green dash-double-dotted line is proportional to $k^{0.7}$.}}
\label{fig2}
\end{figure}
\begin{figure}[!ht]
\centering
\includegraphics[width=8.5cm]{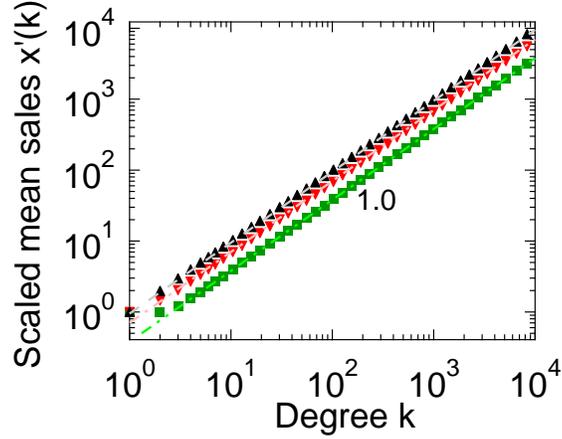}
\caption{\textcolor{black}{{\bf The mean of sales as a function of degree, $x(k)$, for Model-1.} We plot the scaled values [$x'(k)=x(k)/x(1)$].  The black triangles indicate the network for $\gamma=-0.7$, the red nablas for $\gamma=0.0$ and the green squares for $\gamma=0.7$. The mean of sales as a function of degree, $x(k)$, for Model-1.  From this figure, we can confirm that $x(k)$ is proportional to $k$ regardless of $k_{nn}(k)$. These results agree with the approximation given by Eq. \ref{knn1}, namely, $x(k) \propto k$.}}
\label{fig3}
\end{figure}
\begin{figure}[!ht]
\centering
\includegraphics[width=8.5cm]{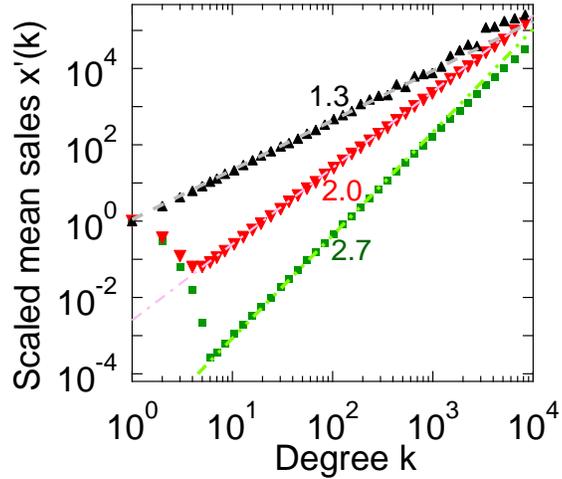}
\caption{\textcolor{black}{{\bf The mean of sales as a function of degree, $x(k)$, for Model-2.} We plot the scaled values [$x'(k)=x(k)/x(1)$]. 
  The black triangles indicate the network for $\gamma=-0.7$, the red nablas for $\gamma=0.0$, the green squares for $\gamma=0.7$, the black dashed line is proportional to $k^2 \cdot k^{-0.7}=k^{1.3}$, the red dash-dotted line is proportional to $k^{2.0}$ and the green dash-double-dotted line is proportional to $k^2 \cdot k^{0.7}=k^{2.7}$. These results agree with the approximation given by Eq. \ref{knn2}; namely, $x(k) \propto k^2 \cdot k_{nn}(k)$. The result for $\gamma=-0.7$ [i.e. $k_{nn}(k) \propto k^{-0.7}$] corresponds to empirical scaling with exponent 1.3.} }
\label{fig4}
\end{figure}

\textcolor{black}{
 \textcolor{black}{Finally}, we \textcolor{black}{numerically} check the approximation by using different artificial networks \textcolor{black}{that} have power-law degree \textcolor{black}{distributions with exponent 1.3 and power-law average degrees of nearest neighbours} like the firm network. }
\textcolor{black}{
The artificial networks \textcolor{black}{satisfy} the following conditions:
\begin{itemize}
\item The degree obeys the power law distribution, $P_k(k) \propto k^{-\alpha-1}$.
\item The average degree of the nearest \textcolor{black}{neighbours} is expressed as a power function of the degree, $k_{nn}(k) \propto k^{\gamma}$.\end{itemize}
To generate networks that satisfy the above conditions, we modify the configuration model as follows:
\begin{enumerate}
\item Assign the degree sequence, $\{k_i\}$, obeying the power law with exponent $-\alpha$ to nodes $\{i\}$. For example, $k_i=floor[(i/M)^{-1/\alpha}]$ ($i=1, 2, \cdots M$), where $M$ is the number of nodes and $floor[x]$ is the largest integer not greater than $x$. 
\item \textcolor{black}{Randomly sample} a node denoted by the $v$ from the set of nodes that have the largest $\{k_i\}$.
\item Update $k_v$; $k_v \leftarrow k_v-1$. 
\item \textcolor{black}{Randomly sample} a node denoted by $w$ from \textcolor{black}{all} nodes except for node $v$ with  \textcolor{black}{probability proportional} to $\{q_i\}$, where $q_i=k_i^{\beta+1}$. 
\item Update $k_w$; $k_w \leftarrow k_w-1$.
\item Connect \textcolor{black}{node $v$} and $w$ with an edge (undirected). 
\item \textcolor{black}{Repeat steps 2} through 6 until $k_i=0$ ($i=1, 2, \cdots M$).
\end{enumerate}
\textcolor{black}{Figures \ref{fig1} and \ref{fig2} show the cumulative distribution the function of degree, which corresponds to $P_k(k)$ and the average degree of nearest neighbours as a function of degree $k_{nn}(k)$ for the artificial network with parameter $\alpha=1.3$ (empirical parameter)\textcolor{black}{,} $\gamma=-0.7$ (empirical parameter), $0.0$ and $0.7$. 
\textcolor{black}{\textcolor{black}{In addition, Figures \ref{fig3}} and \ref{fig4} show} the mean of the sales as a function of degree $x(k)$, which is \textcolor{black}{numerically derived from} Eqs. \ref{model1} and \ref{model2} for the corresponding artificial networks.}
From these figures, we can confirm that all scaling exponents \textcolor{black}{obtained numerically accord with the results} of the approximations given by Eq. \ref{knn1} for Model-1 and by Eq. \ref{knn2} for Model-2. 
Moreover, from \textcolor{black}{Figure \ref{fig4}}, we confirm that only the case $\gamma=-0.7$ , which is the empirical observation of $k_{nn}(k)$ for the firm network, reproduces the empirical scaling 1.3.
}

\section*{Conclusion and discussion}
In this study, by applying the mean-field approximation to the money-transport models given by Eqs. 1 and 2, we were able to consistently understand the relationships, 
 between the \textcolor{black}{power-law} exponents, models and network structures \textcolor{black}{summarised} in Table 1. In particular, we presented the connections of the non-trivial power law scaling with an exponent of 1.3 (in Model-2 for the firm network) with the average degree of the nearest neighbours $k_{nn}(k)$, which is one of the cardinal features of a network.
This result is one of the explanations for empirical scaling relationships with exponent 1.3 between the number of business partners and sales. 
Moreover, \textcolor{black}{non-trivial} empirical scaling with \textcolor{black}{ exponent 1.3} between sales $s$ and number of employees $l$, $s(l) \propto l^{1.3}$, which is observed in Ref. \cite{watanabe2012relations},  might be connected to the network structure. Because, roughly speaking, combining the scaling between sales $s$ and degrees $k$, $s(k) \propto k^{1.3}$, described in this study, and the trivial empirical scaling between number of employees $l$ and degrees $k$, $l(k) \propto k^{1.0}$ reported in Ref. \cite{watanabe2012relations}, we \textcolor{black}{obtain} the scaling between employees and sales, $s(l) \propto l^{1.3}$. 
\section*{Acknowledgments}
The author would like to thank Enago (www.enago.jp) for the English language review.

\bibliography{approx_plos4_br}
\clearpage
\section*{Figures}
\clearpage
\section*{Tables}
\begin{table}[!ht]
\caption{ \bf{Summary of exponents.} }
\centering
\begin{tabular}{lll}
\hline
Model & Network & Exponent $\beta$ \\
\hline
\hline
Equi-partition model & Firm network & 1.0 \\
& Shuffled network & 1.0 \\
Weighted partition model & Firm network&  1.3 \\
& Shuffled network & 2.0 \\
\hline
Sales (empirical data) & Firm network & 1.3 \\
\hline
\end{tabular}
\label{t1}
\end{table}
\end{document}